\documentclass[11pt]{article}

\usepackage{latexsym,amsmath,amssymb,amscd,eucal}
\usepackage{xypic}
\usepackage{enumerate}

\def\harr#1#2{\smash{\mathop{\hbox to .3in{\rightarrowfill}}
 \limits^{\scriptstyle#1}_{\scriptstyle#2}}}

\def\appendix#1{\addtocounter{section}{1}\setcounter{equation}{0}
\renewcommand{\thesection}{\Alph{section}}
\section*{Appendix \thesection\protect\indent \parbox[t]{11.715cm} {#1}}
\addcontentsline{toc}{section}{Appendix \thesection\ \ \ #1} }
\newcommand{\eq}{\begin{equation}}
\newcommand{\eqend}{\end{equation}}

\newbox\ncintdbox \newbox\ncinttbox
\setbox0=\hbox{$-$} \setbox2=\hbox{$\displaystyle\int$}
\setbox\ncintdbox=\hbox{\rlap{\hbox to
\wd2{\hskip-.125em\box2\relax \hfil}}\box0\kern.1em}
\setbox0=\hbox{$\vcenter{\hrule width 4pt}$}
\setbox2=\hbox{$\textstyle\int$}
\setbox\ncinttbox=\hbox{\rlap{\hbox to
\wd2{\hskip-.175em\box2\relax \hfil}}\box0\kern.1em}

\def\be{\begin{equation}}
\def\ee{\end{equation}}
\def\bea{\begin{eqnarray}}
\def\eea{\end{eqnarray}}
\def\bd{\begin{displaymath}}
\def\ed{\end{displaymath}}

\DeclareFontFamily{U}{rsf}{}
\DeclareFontShape{U}{rsf}{m}{n}{
  <5> <6> rsfs5 <7> <8> <9> rsfs7 <10-> rsfs10}{}
\DeclareMathAlphabet\Scr{U}{rsf}{m}{n}

\def\cR{{\Scr R}}

\def\cB{{\Scr B}}
\def\cC{{\Scr C}}

\def\cH{{\Scr H}}

\def\cE{{\Scr E}}

\makeatletter
\newdimen\normalarrayskip              % skip between lines
\newdimen\minarrayskip                 % minimal skip between lines
\normalarrayskip\baselineskip
\minarrayskip\jot
\newif\ifold             \oldtrue            
\def\arraymode{\ifold\relax\else\displaystyle\fi} % mode of array entries
\def\@arrayskip{\ifold\baselineskip\z@\lineskip\z@
     \else
     \baselineskip\minarrayskip\lineskip2\minarrayskip\fi}
\def\@arrayclassz{\ifcase \@lastchclass \@acolampacol \or
\@ampacol \or \or \or \@addamp \or
   \@acolampacol \or \@firstampfalse \@acol \fi
\edef\@preamble{\@preamble
  \ifcase \@chnum
     \hfil$\relax\arraymode\@sharp$\hfil
     \or $\relax\arraymode\@sharp$\hfil
     \or \hfil$\relax\arraymode\@sharp$\fi}}
\def\@array[#1]#2{\setbox\@arstrutbox=\hbox{\vrule
     height\arraystretch \ht\strutbox
     depth\arraystretch \dp\strutbox
     width\z@}\@mkpream{#2}\edef\@preamble{\halign \noexpand\@halignto
\bgroup \tabskip\z@ \@arstrut \@preamble \tabskip\z@ \cr}%
\let\@startpbox\@@startpbox \let\@endpbox\@@endpbox
  \if #1t\vtop \else \if#1b\vbox \else \vcenter \fi\fi
  \bgroup \let\par\relax
  \let\@sharp##\let\protect\relax
  \@arrayskip\@preamble}
\makeatother

\newcommand{\beq}{\begin{eqnarray}}
\newcommand{\eeq}{\end{eqnarray}}

\def\appendix#1{\addtocounter{section}{1}\setcounter{equation}{0}
\renewcommand{\thesection}{\Alph{section}}
\section*{Appendix \thesection. #1}
\addcontentsline{toc}{section}{Appendix \thesection\ \ \ #1} }

\numberwithin{equation}{section}

\begin{document}

%\begin{flushright}
%SISSA/nn/06/EP\\ hep-th/0602162n
%\end{flushright}

\vspace{.1in}
%Preliminary Version

\begin{center}

{\Large\bf GLOBAL ANOMALY AND A FAMILY OF STRUCTURES ON FOLD PRODUCT OF COMPLEX TWO-CYCLES}

\end{center}
\vspace{0.1in}
\begin{center}
{\large
A. A. Bytsenko %$^{(a)}$
\footnote{abyts@uel.br}}
%and M. E. X. Guimar\~aes $^{(c)}$ %\footnote{emilia@if.uff.br}
\vspace{7mm}
\\
{\it Departamento de F\'{\i}sica, Universidade Estadual %de
Londrina\\
Caixa Postal 6001, Londrina-Paran\'a, Brazil}
\end{center}
\vspace{0.1in}
%~\\
\begin{center}
{\bf Abstract}
\end{center}

We propose a new set of IIB type and eleven-dimensional supergravity
solutions which consists of the $n$-fold product of two-spaces ${\bf H}^n/\Gamma$ (where ${\bf H}^n$ denotes the product of $n$ upper half-planes $H^2$ equipped with the co-compact action of $\Gamma \subset SL(2, {\mathbb R})^n$) and $({\bf H}^n)^*/\Gamma$ (where $(H^2)^* = H^2\cup \{{\rm cusp\,\, of}\,\,\Gamma\}$ and $\Gamma$ is a congruence subgroup of $SL(2, {\mathbb R})^n$). The Freed-Witten global anomaly condition have been analyzed.  We argue that the torsion part of the cuspidal cohomology involves in the global anomaly condition.  Infinitisimal deformations of generalized complex (and K\"ahler) structures also has been analyzed and stability theorem in the case of a discrete subgroup of $SL(2, {\mathbb R})^n$ with the compact quotient ${\bf H}^n/\Gamma$ was verified.

\newpage

\section{Introduction}

New infinite classes of eleven-dimensional supergravity warped product
of AdS$_3$ with an eight-dimensional manifold $X_8$, which are dual to
conformal field theories with $N=(0,2)$ supersymmetry (since of the
AdS-CFT correspondence), have been found in \cite{Waldram1,Waldram2}.
These new solutions are all $S^2$ bundles over six-dimensional base
spaces. A two-sphere bundle can be received from the canonical
line-bundle over base spaces $\cB_6$ by adding a point at infinity to
each of the fibers (see for detail \cite{Waldram1}). Spaces $\cB_6$
are products of K\"ahler-Einstein $n$-spaces ($KE_n$) with various
possibilities for the signs of the curvature:
\begin{itemize} 
\item{} The class of the most general polynomial solution 
$\cB_6=KE_2\times KE_2\times KE_2$.
\item{} The class solutions for which $\cB_6=KE_4\times KE_2$ and
  $\cB_6=KE_6$; these solutions can be obtained from the general
  solution as special cases.
\end{itemize} 

A general class of solutions of K\"ahler-Einstein spaces of the form
$X_{2n} = KE_2^{(1)}\times \cdots \times KE_2^{(n)}$, where $KE_2$ is
a two-dimensional space of negative curvature, is our special
interest.  In this note we concentrate on solutions which are the
$n$-fold product of two-spaces ${\bf H}^n/\Gamma$ (where ${\bf H}^n$
denotes the product of $n$ upper half-planes $H^2$ equipped with the
co-compact action of $\Gamma \subset SL(2, {\mathbb R})^n$) and $({\bf
  H}^n)^*/\Gamma$ (where $(H^2)^* = H^2\cup \{{\rm cusp\,\,
  of}\,\,\Gamma\}$ and $\Gamma$ is a congruence subgroup of $SL(2,
{\mathbb R}^n)$). Some homological and K-theory methods applied for
hyperbolic cycles the reader can find in
\cite{BytsenkoPoS,Bonora,Bytsenko}.  We analyze the Freed-Witten
global anomaly condition and argue that the torsion part of the
cuspidal cohomology involves in that condition. Also we verify the
stability theorem for a discrete subgroup of $SL(2, {\mathbb R})^n$
with compact quotient ${\bf H}^n/\Gamma$.

\section{The Hilbert modular groups and varieties}
\label{Hilbert0}

In this section we consider some examples of general factorized solutions. This class includes the product of $n$-fold two-space forms which associated with discrete subgroups of $SL(2, {\mathbb R})^n$ with compact quotient ${\bf H}^n/\Gamma$, the Hilbert modular groups and varieties. We will mostly followed \cite{Freitag} in reproducing of necessary results. The central topics are the following:
\begin{itemize}
\item{} A discrete subgroup $\Gamma\subset SL(2, {\mathbb R})^n$ with compact quotient ${\bf H}^n/\Gamma$ and the Hilbert modular groups. (The corresponding singular cohomology groups $H^\bullet (\Gamma; {\mathbb C})$ has beed determed in 
\cite{Matsushima}.)
\item{} The Elenberg-MacLane cohomology groups 
$ H^{\bullet}(\Gamma_{\bf k}; {\mathbb C})$,
where $\Gamma_{\bf k}$ acts trivially on $\mathbb C$ and totally real number field ${\bf k}\supset {\mathbb Q}$. The corresponding spaces 
are the {\it Hilbert modular varieties}:
${\bf H}^n/\Gamma_{\bf k}   \equiv (\underbrace{
{H}^2\times \ldots \times{H}^2}_n)/\Gamma_{\bf k}$.
\item{} The mixed Hodge structure in the sence of Deligne \cite{Deligne}.
\end{itemize}
The Hilbert modular group $\Gamma_{\bf k}=SO(2, {\bf o})$, the corresponding spaces (Hilbert modular varieties) and functions ({\it Hilbert modular forms}) have been actively studied in mathematics (for the reference see \cite{Freitag}).
$\Gamma_{\bf k}$ is the group of all $2\times 2$ matricies of determinant 1 with coefficients in the ring $\bf o$ of integers of a totally real number field. 
The Eilenberg-MacLane cohomology groups
$H^\bullet(\Gamma_{\bf k}; {\mathbb C})$ are isomorphic to the singular cohomology group of Hilbert modular variety 
$
H^\bullet (\Gamma_{\bf k}; {\mathbb C}) = 
H^\bullet ({\bf H}^n/\Gamma_{\bf k};
{\mathbb C}),
$
where as before ${\bf H}^n$ denotes the product of n upper half-planes $H^2$ equipped with the natural action of $\Gamma_{\bf k}$. The Hilbert modular variety carries a natural structure as a {\it quasiprojective variety} and its cohomology groups inherit a {\it Hodge structure}. In fact the Hilbert modular group is a simplified example of the cohomology theory of arithmetic groups
and it is the only special case in which the cohomology can be determined explicitly.

\subsection{The global anomaly condition}
Let us consider a bundle whose fiber $\xi$ is $(p-1)$-connected;
this means that for $k<p$ the $k$-th homotopy group 
$\pi_{k< p}(\xi)$ of $\xi$ vanishes. Let the $p$th homotopy group ${G}$ be non-trivial and $\pi_{k=p}(\xi) = {G}$. Then the $\xi$ bundle can be specified by a degree $(p+1)$ characteristic class in the cohomology with coefficients in 
$G$, $\omega_{p+1}\in H^{p+1}(X; {G})$, and some characteristic classes of higher degree. If all of the homotopy classes of $\xi$ of degree higher than $p$ vanish, then the bundle is characterized by $\omega_{p+1}$. 

If ${G} = {\mathbb Z}_2$ and $p=0$ then a non-vanishing characteristic class $\omega_1\in H^1(X; {\mathbb Z}_2)$ is associated with existence of the so-called {\it spin} structure on $X$. For the circle bundle, $p=1,\, {G}= {\mathbb Z}$, we have a single characteristic class $c_1=\omega_2\in H^2(X; {\mathbb Z})$ which is called {\it the first Chern class}. 
For every degree $(p+1)$ characteristic class there is a degree $(p+2)$ obstruction to a lift. The obstruction to the existence of a spin structure is the third Stiefel-Whitney class $\omega_3\in H^3(X; {\mathbb Z})$. The class $\omega_3$ is always ${\mathbb Z}_2$ torsion. Unlike $\omega_2$ the class $\omega_3$ always has lift to cohomology with integral coefficients, which is denoted  $W_3\in H^3(X; {\mathbb Z})$ and also always ${\mathbb Z}_2$ torsion. $W_3$ is defined to be the Bockstein homomorphism $\beta$ of $\omega_2$, $W_3= \beta\omega_2$. If the third Stiefel-Whitney class of the tangent bundle $TX$ is equal zero then $X$ is said to be spin manifold (and a spin lift of the tangent bundle exists).

Quantum field theory on a fold product of compact spaces has to be anomaly free. It means that the Freed-Witten anomaly condition must be holds
\begin{equation}
W_3(X) + [H]\vert_{X} =0 \,\,\,\, {\rm in }
\,\,\,\, H^3(X;\mathbb Z)\,.
\label{condition}
\end{equation} 
Here $H$ is the pullback of the NSNS three-form to the space worldvolume $X$. 
\footnote{
Note that in the de Rham theory
$[H]_{DR}\vert_{X}=0$. In the bosonic string we must impose condition (\ref{condition}) without the $W_3(X)$ term.
}
Since branes can wrap cycles we discuss briefly the condition  (\ref{condition}) for branes in string theory. It is known that a brane wrapping homologically nontrivial cycle $Y$ can
nevertheless be unstable if for some $Y'\subset X$
the following condition holds \cite{Maldacena}:
\begin{equation}
PD(Y \subset Y') = W_3(Y') + [H]\vert_{Y'}\,.
\end{equation}
Here the left hand side denotes the Poincare dual of $Y$ in
$Y'$. (In the bosonic string the question of stability is more complicated because they always include tachyons.) Free brane can wrap any homologically nontrivial cycle in $X$. Also a brane wrapping a nontrivial cycle is absolutely stable \cite{Maldacena}.
It should be noted that branes with the Freed-Witten anomalies cannot carry K-theory charges and they are inconsistent. A brane which wraps a non-singular submanifold and anomaly-free carries a K-theory charge. D-branes can fails to carry K-theory charge if: it is the Freed-Witten anomalous or it wraps homology cycles which cannot be represented by any non-singular submanifold.

\noindent
\\
{\bf The Hilbert modular varieties.} 
The differential $d_3$ in K-theory has the form \cite{Atiyah,Rosenberg}
\begin{equation}
d_3 = Sq^3,\,\,\,\,\,\, d_3 = Sq^3 + [H]\,\,\,\,\,\,{\rm for\,\,\, twisted\,\,\, K-theory}\,,
\end{equation}
where the Steenrod square $Sq^3$ takes an integral class in the $k$th cohomology to a class in the $(k+3)$rd cohomology as does cup product with $[H]$.
A necessary (but not sufficient) condition for the vanishing of $(W_3(X) + [H]|_X)$ on a worldvolume is: 
the flux has to be in the kernel of the spectral sequence differential
$
d_3 = Sq^3 + [H] \cup.
$
In the de Rham theory, for example, we obtain the simple expression
$
d_3(\omega) = [H]\wedge \omega\,.
$
Unlike the cup product with $[H]$, $Sq^3$ is only nontrivial when acting on ${\mathbb Z}_2$ torsion components of $H^k(X)$ and the image is likewise always a ${\mathbb Z}_2$ torsion component of $H^{k+3}(X)$.
Generally speaking in passing from $K(X)$ to the full K-theory group one needs to solve an extension problem to obtain the
correct torsion subgroup.

Analysing the global anomaly condition recall that for any $X$, $m$, and associated abelian group $G$ the following result holds (the universal coefficient theorem):
the homology and cohomology group of $X$ with coefficients in $G$ has a splitting 
\begin{eqnarray}
H_m(X; G) & \cong & H_m(X)\otimes G 
\oplus {\rm Tor}\, (H_{m-1}(X; G))\,,
\nonumber \\
H^m(X; G) & \cong & H^m(X)\otimes G 
\oplus {\rm Tor}\, (H^{m+1}(X; G))\,,
\nonumber \\
H^m(X; G) & \cong &
{\rm Hom}\,(H_m(X); G)\oplus {\rm Ext}\,
(H_{m-1}(X); G)\,.
\end{eqnarray}
Here $H^m(X)$\,\,($H_m(X))$ are the cohomology (homology) groups with integer coefficients.
The (splittings) isomorphisms given by the universal coefficient theorem are said to be unnatural isomorphisms. The following maps of exact sequences are natural:
\begin{equation}
\begin{array}{ccccccccc}
0 & \longrightarrow & \!\!\!\!\!\!\!\!\!\!\!\!\! H_m(X)\otimes G  & \longrightarrow & H_m(X; G) & \longrightarrow & 
\!\!{\rm Tor}(H_{m-1}(X); G)  & \longrightarrow & 0\,
\\
0 & \longrightarrow & \!\!\!\!\!\!\! H^m(X; {\mathbb Z})\otimes G   & \longrightarrow & H^m(X; G)  & \longrightarrow & 
\,\,\,\,\,{\rm Tor}(H^{m+1}(X; {\mathbb Z}), G) & \longrightarrow & 0\,
\\
0 & \longleftarrow & {\rm Hom}(H_m(X); G) & \longleftarrow & H^m(X; G)
& \longleftarrow & {\rm Ext}(H_{m+1}(X); G) & \longleftarrow & 0\,
\end{array}
\label{sec}
\end{equation}
For example, the first exact sequence of (\ref{sec}) can be deduced
as follows. Let $G={\mathfrak G}_1/{\mathfrak G}_2$, where ${\mathfrak G}_1$ and ${\mathfrak G}_2$ are the abelian groups. It is clear that 
$H_m(X; {\mathfrak G}_j)= H_m(X)\otimes {\mathfrak G}_j$, where the group 
${\mathfrak G}_j$ is a sum of set of groups $\mathbb Z$. Thus,
$
H_m(X; {\mathfrak G}_j) = H_m(X; {\mathbb Z}\oplus {\mathbb Z}\oplus 
\cdots ) = H_m(X)\oplus H_m(X)\oplus \cdots = H_m(X)\otimes
{\mathfrak G}_j.
$
Let us consider the fifth-term fragment of a sequence
\begin{eqnarray}
&&
\!\!\!\!\!\!\!\!\!\!\!\!\!\!\!\!
H_m(X; {\mathfrak G}_2)\,\,\,\,\longrightarrow 
H_m(X; {\mathfrak G}_1)\,\,\,\,\longrightarrow
H_m(X; G)\longrightarrow
H_{m-1}(X; {\mathfrak G}_2)\,\,\,\longrightarrow
H_{m-1}(X; {\mathfrak G}_1) \,\,\,\,\,\, \Longrightarrow
\nonumber \\
&&
\!\!\!\!\!\!\!\!\!\!\!\!\!\!\!\!
H_m(X)\otimes {\mathfrak G}_2\longrightarrow
H_m(X)\otimes {\mathfrak G}_1\longrightarrow
H_m(X; G)\longrightarrow
H_{m-1}(X)\otimes {\mathfrak G}_2\longrightarrow
H_m(X)\otimes {\mathfrak G}_1
\label{sec1}
\end{eqnarray}
Note that any fifth-term exact sequence
$
A\stackrel{\sigma}{\rightarrow} B\rightarrow C\rightarrow D
\stackrel{\tau}{\rightarrow} E
$
can be transformed to a short exact sequence
$
0\rightarrow {\rm Coker}\,\sigma\rightarrow C
\rightarrow {\rm Ker}\,\tau\rightarrow 0,
$
where ${\rm Coker}\,\sigma = B/{\rm Im}\,\sigma.
$
This transform converts the last fifth-term sequence to the first short sequence in (\ref{sec}).
The second sequence of (\ref{sec}) can be similary deduced, while the varification of the last sequence is requied more complicate arguments.

The bifiltered spectral sequence of Hilbert modular varieties really computes the associated graded space. 
Note that for strictly co-compact subgroup $\Gamma \subset SL(2, {\mathbb R})^n$ the group of cohomology $H^{m=3}(\Gamma)$ vanishes. From the second sequence of (\ref{sec}) it follows that for $m=3$ and 
$G= {\mathbb C}$ the group 
$H^3(X={\bf H}^n/\Gamma; {\mathbb Z})\otimes {\mathbb C}$ vanishes.
Therefore we conclude that in the case of co-compact group $\Gamma$ the torsion-free group of cohomology 
$H^3({\bf H}^n/\Gamma; {\mathbb Z})$ is trivial (the class of $[H]_X$ is trivial).

More complicate situation occurs is in the presence of the 
cuspidal part of the total group of cohomology. For a discrete subgroup $\Gamma_\kappa \subset SL(2, {\mathbb R})^n$ and each boundary point $\kappa \in {\mathbb R}\cup\{\infty\}$ there exist the parabolic elements in the stabilizer $\Gamma_\kappa$. 
All class of cohomologies can be maped to a boundary of the space. Therefore the cuspidal contribution 
$
H^{n}_{\rm cusp}(\Gamma)
$ 
(see \cite{Freitag} for details) could be involves in the global anomaly condition (\ref{condition}).

Finally note that for rigorous analysis of the global anomaly 
one plainly needs to have an explicit expression for the higher differentials $\{d_\ell\}_{\ell=3}^{2n-1}$. An expression for the differential $d_3$ is known, but not much is known about the higher differentials in general. Here we restrict ourselves to the case of $d_3$ only.

\section{Stability of generalized complex (and K\"ahler) structures}
\label{stability}

{\bf Differential Gerstenhaber algebra and infinitisimal deformations.}
Let us consider a ring $\cR$ with unit and let $A$ be an 
$\cR$-algebra. Suppose that ${\mathfrak a} =\oplus_{n\in {\mathbb Z}}{\mathfrak a}^n$ is a graded algebra over $A$. If $a\in {\mathfrak a}^n$, let $|a|$ denote its degree. 
The algebra ${\mathfrak a}$ is called a Gerstenhaber algebra if there is an associative product $\wedge$ and a graded commutative product $[-\bullet -]$ 
satisfying the following axioms: for $a\in {\mathfrak a}^{|a|}$, $b\in {\mathfrak a}^{|b|}$,  $c\in {\mathfrak a}^{|c|}$\,,
\begin{eqnarray}
\!\!\!\!\!\!\!\!\!
&& a\wedge b \in {\mathfrak a}^{|a|+ |b|},\,\,\,\,\,\,\, b\wedge a = 
(-1)^{|a||b|}a\wedge b\,,
\nonumber \\
\!\!\!\!\!\!\!\!\!
&& (-1)^{(|a|+1)(|c|+1)}[[a\bullet b]\bullet c ] \, + \, 
{\rm more}\,\,\,{\rm two}\,\,\,\,{\rm terms}\,\,\,\,(a, b, c \,\,\,{\rm cyclic}\,\,\,{\rm permutation}) = 0\,, 
\nonumber \\
\!\!\!\!\!\!\!\!\!
&& [a\bullet b\wedge c] = [a\bullet b]\wedge c + (-1)^{(|a|=1)|b|}
b\wedge [a\bullet c]\,.
\end{eqnarray}
Note that a differential graded algebra ${\mathfrak a}$ is a graded algebra with a graded commutative product $\wedge$ and a differential $d$ of degree +1, i.e. a map $d: {\mathfrak a}\rightarrow  {\mathfrak a}$ such that
\begin{equation}
d({\mathfrak a}^n)\subseteq {\mathfrak a}^{n+1},\,\,\,\, 
d\circ d = 0,\,\,\,\, 
d(a\wedge b)= da\wedge b + (-1)^{|a|}a\wedge db\,.
\end{equation}
Let ${\mathfrak a}$ be a graded algebra over $\mathfrak C$ such that $({\mathfrak a}, [-\bullet -], \wedge)$
form a Gerstenhaber algebra and $({\mathfrak a}, \wedge, d)$ form a differential graded algebra. If in addition
\begin{equation}
d[a\bullet b] = [ad\bullet b] + (-1)^{|a|+1}[a\bullet db]\,,
\,\,\,\,\,\forall\, a,b \in {\mathfrak a}\,,
\end{equation}
then $({\mathfrak a}, [-\bullet -], \wedge, d)$ is a differential Gerstenhaber algebra.
Let $\mathbb J$ be an integrable complex structure on a Lie algebra $\mathfrak g$, i.e. $\mathbb J$ is an endomorphism
of $\mathfrak g$ such that ${\mathbb J}^2 = -$id and
\begin{equation}
[x\bullet y] + {\mathbb J}[{\mathbb J}x\bullet y] + 
{\mathbb J}[x\bullet {\mathbb J}y] -
[{\mathbb J}x\bullet {\mathbb J}y] = 0\,.
\end{equation}
Therefore the $\pm i$ eigenspaces ${\mathfrak g}$ and ${\mathfrak g}^{*}$ are complex Lie subalgebras of the complexified algebra
${\mathfrak g}^{\mathbb C}$.
Note that if $({\mathfrak g}, [- \bullet -])$ is a Lie algebra, the Chevalley-Eilenberg differential $d$ is defined on the dual vector space ${\mathfrak g}^*$ by the relation $d\alpha (x,y):= -\alpha([x\bullet y])$, for $\alpha\in {\mathfrak g}^*$ and $x, y \in {\mathfrak g}$. The identity $d\circ d=0$ is equivalent to the Jacobi identity for the Lie bracket $[-\bullet -]$ on $\mathfrak g$. It follows that $(\wedge{\mathfrak g}^*, d)$ is a differential graded algebra.

Let $X$ be a compact K\"ahler manifold with the complex structure $J$ and the K\"ahler form $\omega$.
We are looking for the generalized K\"ahler structure 
$({\mathbb J}, e^{\sqrt{-1}\omega})$.
Let us consider the decomposition
$
({\mathfrak g}\oplus {\mathfrak g}^*)\otimes {\mathbb C}:=(T\oplus T^*)\otimes {\mathbb C} = L\oplus \overline{L}\,.
$
The differential Gerstenhaber algebra ${\mathfrak E} = (\Omega^\bullet \overline{L}, d_{\overline{L}})$ is elliptic and it gives rise to a Kuranishi deformation theory for any generalized complex structure.
Let $\{{\mathfrak U}_\alpha\}$ be an open cover of $X$ and $\Omega_\alpha$ a nowhere vanishing holomorphic $n$-form on ${\mathfrak U}_\alpha$.
It has been shown \cite{Goto} that there is the isomorphism between complexes:
\begin{equation}
(\Omega^\bullet \overline{L},\,d_{\overline{L}}) \cong 
(U^{-n+\bullet}\otimes {\cC}^{-1}_J,\, \overline{\partial})\,,
\end{equation}
where ${\cC}^{-1}_J$ denotes the dual of the standard canonical line bundle of the complex manifold $(X, J)$. 
The space $U^{-n} := {\cC}_{\mathbb J}$ is a complex line bundle which we call the canonical line bundle of $\mathbb J$.
If $\wedge^k\overline{L}_{\mathbb J}$ is the $k$-th exterior product of $\overline{L}_{\mathbb J}$,
then the eigenspace $U^{k}$ is given by the action of $\Omega^k\overline{L}_{\mathbb J}$ on 
${\cC}_{\mathbb J}$,\,
$
U^{-n+k}=\Omega^k\overline{L}_{\mathbb J} \cdot 
{\cC}_{\mathbb J}.
$
The space of infinitesimal deformations of complex structures on $X$ is given by the direct sum of the 
${\cC}^{-1}_{J}$-valued Dolbeault cohomology groups
\begin{equation} \label{deformations}
{\cH}^{n,2}_{\overline{\partial}}(X; 
{\cC}^{-1}_{J})\oplus 
{\cH}^{n-1,1}_{\overline{\partial}}(X; 
{\cC}^{-1}_{J})\oplus 
{\cH}^{n-2,0}_{\overline{\partial}}(X; 
{\cC}^{-1}_{J}),
\end{equation}
where we marked cohomology groups by the operator $\overline{\partial}$.
The spaces of infinitisimal deformations  of complex structures are:

The space  ${\cH}^{n,2}_{\overline{\partial}}(X; {\cC}^{-1}_{J})$ is given by the action of {B}-fields {\rm (}$2$-forms{\rm )}\,.

The space ${\cH}^{n-2,0}_{\overline{\partial}}(X; {\cC}^{-1}_{J})$ is induced by the action of holomorphic $2$-vector fields\,.

The space of the obstructions is given by
\begin{equation} \label{obstructions} {\cH}^{n,3}_{\overline{\partial}}(X; {\cC}^{-1}_{J})\oplus
{\cH}^{n-1,2}_{\overline{\partial}}(X; {\cC}^{-1}_{J})\oplus {\cH}^{n-2,1}_{\overline{\partial}}(X;
{\cC}^{-1}_{J})\oplus {\cH}^{n-3,0}_{\overline{\partial}}(X; {\cC}^{-1}_{J})\,.
\end{equation}
As it follows from Eqs. (\ref{deformations}) and (\ref{obstructions})
the space ${\cH}^{n-1,1}_{\overline{\partial}}(X; {\cC}^{-1}_{J})\cong H^1(X;\Theta)$ is the space of infinitesimal
deformations of complex structures in Kodaira-Spencer theory.

\noindent
\\
{\bf The theorem of stability} (\cite{Goto}, Theorem 5.1). \label{ST}
The main result can be formulated as the theorem of stability. Suppose
that $X$ is a compact K\"ahler manifold with the K\"ahler structure
$({\mathbb J}, e^{\sqrt{-1}\omega})$. If the obstruction space $
\oplus_{i=0}^3 {\cH}^{n-i,3-i}_{\overline{\partial}}(X;
{\cC}^{-1}_{J}) $ vanishes, then we have the family of
K\"ahler structures $\{{\mathbb J}_t, \psi_{t,s}\}$ with $({\mathbb J}_0, \psi_{ 0,0})=({\mathbb J}, e^{\sqrt{-1}\omega})$ which is parametrized by $(t,s) \in W\times W'$, where $W$ is a small open set of $ \oplus_{i=0}^2 {\cH}^{n-i, 2-i}_{\overline{\partial}}(X;
{\cC}^{-1}_{J}) $ and $W'$ denotes a small open set of
${\cH}^{1,1}_{\overline{\partial}}(X)$ containing the origin.

\subsection{The compact quotient ${\bf H}^n/\Gamma$}

For deforming of a complex manifold $(X, J)$ as a generalized complex
manifold let us consider a subspace $T\oplus T^*$.  The deformation
complex becoming $ \oplus_{p+q=m}\Lambda^{0, q}(\Omega^pT^*, \,
\overline{\partial}), $ where $\Lambda^{0, q}$ is a generator of the
$(0, q)$-forms for a complex $n$-dimensional space.  Recall that
$\mathbb J$ determines an alternating grading for the differential
forms, and the integrability of $\mathbb J$ is equivalent to the fact
that the exterior derivative $d$ splits into the sum $d= \partial +
\overline{\partial}$.  The elliptic complex $({\cC}^\bullet,
d)$ is a subcomplex of the (full) de Rham complex $ \cdots \overset
d\to \Omega^\bullet T^*X\overset d\to \cdots $ and the cohomology
group of the de Rham complex is given by the de Rham cohomology $
H(X):= \oplus_i H^i(X; \mathbb C).  $ A $(p, q)$ decomposition for the
de Rham cohomology of any compact generalized K\"ahler manifold is $
H^\bullet (X; {\mathbb C}) = \oplus_{p,q}
{\cH}^{p, q}\,, $ where ${\cH}^{p,q}$ are harmonic forms
in $U^{p,q}$.  In general for the compact quotient $X\equiv {\bf
H}^n/\Gamma$ ($\Gamma$ has no elliptic fixed points) one has $
{\cH}^{p, q}(\Gamma)\cong H^q({\bf H}^n/\Gamma; {\cE}^p), $ where ${\cE}^p$ denotes the sheaf of holomorphic
$p$-forms on the analytic manifold ${\bf H}^n/\Gamma$.

For the illustration of stability theorem let us choose $n=2$. There
is the induced map $ p^i_{\cC}:\, H^i({\cC}^\bullet)\rightarrow H(X) $ (de Rham cohomology), which is
injective for $i=1,2$ \cite{Goto}.  For any $n$ and $p=q$ the
universal part of the Hodge cohomology ${\cH}^{p, q}_{\rm
univ}(\Gamma)$ is non-zero \cite{Freitag}, and for $p=q=\ell$ there
is a set of groups: $\{{\cH}^{\ell, \,\ell}_{\rm
univ}(\Gamma)\}_{\ell=0}^n$.  It is clear that if $n=2$ then the
obstruction space
%$
%\oplus_{i=0}^3 \cH^{n-i,3-i}_{\overline{\partial}}
%(X, \cC^{*}_{J})
%$ 
vanishes and the universal part of cohomology contributes to the space
of infinitisimal deformations.  From the theorem of stability it
follows that there is a family of generalized K\"ahler structures on
${\bf H}^2/\Gamma$ which is parametrized by an open set of the direct
sum $ \oplus_{i=0}^2 {\cH}^{2-i, 2-i}_{\overline{\partial}}(X;
{\cC}^{-1}_{J}).  $

There is also a family of deformations of generalized complex
structures on ${\bf H}^n/\Gamma$ which can be parametrized by the
space of holomorphic $n$-vector fields $H^0({\bf H}^n/\Gamma,
\Omega^n_h)$. Indeed, let $\{{\mathfrak T}_\alpha, \Omega_\alpha\}$ be
a trivilization of the canonical line bundle $\mathfrak C$, and let
$V$ be a holomorphic $n$-vector fiend on $X$. One can take coordinates
$\{dz_a^\alpha\}$ on each ${\mathfrak T}_\alpha$ and
\begin{equation}
\Omega_\alpha = dz_{j_1}^{\alpha}\wedge\cdots \wedge dz_{j_n}^{\alpha},
\,\,\,\,\,\,\,\,\,
V = f_{j_1,\ldots,j_n}\frac{\partial}{\partial z_{j_1}^\alpha}
\wedge \cdots \wedge \frac{\partial}{\partial z_{j_n}^\alpha}\,.
\end{equation}
It follows $ \exp (V)\wedge \Omega_{\alpha} = f_{j_1,\ldots,j_n} +
\Omega_{\alpha}\,.  $ Therefore $\exp (V)\wedge \Omega_{\alpha}$ is a
non-degenerate spinor which induces a generalized complex structure
${\mathbb J}_{V}$. It should be stressed that the type of
generalized complex structure $\mathbb J$ is defined as the minimal
degree of differential forms (non-degenerate pure spinors) which
induces ${\mathbb J}$.  One can find that the first cohomology of the
complex $({\cC}^\bullet, d)$ is described as $ H^1({\cC}^\bullet)\cong {\cH}^{1,1}_{\overline{\partial}} ({\bf
H}^n/\Gamma).  $ Hence from the theorem of stability we have a
family of K\"ahler structures on ${\bf H}^n/\Gamma$ parametrized by
$H^0({\bf H}^n/\Gamma; \Omega_h^{n})\oplus {\cH}^{1,1}_{\overline{\partial}} ({\bf H}^n/\Gamma)$.

\end{document}